\documentclass{iopart}

\usepackage{iopams}
\usepackage{graphicx}  
\usepackage{subfigure}

\begin{document}

\title{Proximity effect in superconductor/conical magnet/ferromagnet
  heterostructures}

\author{Daniel Fritsch and James F. Annett}

\address{H. H. Wills Physics Laboratory, School of Physics, University
  of Bristol, Bristol BS8 1TL, UK}

\ead{daniel.fritsch@bristol.ac.uk, james.annett@bristol.ac.uk}

\begin{abstract}
At the interface between a superconductor and a ferromagnetic metal
spin-singlet Cooper pairs can penetrate into the ferromagnetic part of
the heterostructure with an oscillating and decaying spin-singlet
Cooper pair density. However, if the interface allows for a
spin-mixing effect, equal-spin spin-triplet Cooper pairs can be
generated that can penetrate much further into the ferromagnetic part
of the heterostructure, known as the long-range proximity
effect. Here, we present results of spin-mixing based on
self-consistent solutions of the microscopic Bogoliubov-de Gennes
equations incorporating a tight-binding model. In particular, we
include a conical magnet into our model heterostructure to generate
the spin-triplet Cooper pairs and analyse the influence of conical and
ferromagnetic layer thickness on the unequal-spin and equal-spin
spin-triplet pairing correlations. It will be show that, in agreement
with experimental observations, a minimum thickness of the conical
magnet is necessary to generate a sufficient amount of equal-spin
spin-triplet Cooper pairs allowing for the long-range proximity
effect.
\end{abstract}

\pacs{74.45.+c, 74.78.Fk, 74.20.-z, 74.20.Mn}

\submitto{\NJP}

\section{Introduction}
\label{Introduction}

At the interface between a normal metal and a superconductor (SC) an
incoming electron with an energy above the Fermi energy (or chemical
potential) $\mu$ can be reflected back into the metal as a hole with
opposite spin orientation. This phenomenon, known as Andreev
reflection~\cite{Andreev_JETP19_1228}, gives rise to the well-known
proximity effect resulting in superconducting properties decaying into
the normal metal part of the heterostructure. Replacing the normal
metal by a ferromagnet (FM) drastically changes the behaviour at the
interface. At first sight the phenomenona of ferromagnetism and
superconductivity appear to be mutually exclusive due to the specific
requirements concerning spin orientations. In ferromagnetic materials
the Pauli principle requires the spins to orient parallel whereas
superconducting spin-singlet Cooper pairs require an antiparallel
orientation. Based on these intrinsic spin orientations there are
interesting phenomena to be expected at the interface between a
ferromagnet and a superconductor. The exchange interaction in the
ferromagnet leads to different Fermi velocities of electrons in the
spin-up and spin-down channel. Therefore, the centre of mass motion is
modulated and superconducting correlations in the ferromagnet show an
oscillating
behaviour~\cite{Buzdin_JETPLett35_178,Buzdin_JETPLett53_321,Demler_PRB55_15174}. These
are essentially the FFLO oscillations named after Fulde and
Ferrell~\cite{Fulde_PhysRev135_A550} and Larkin and
Ochinikov~\cite{Larkin_JETP20_762}, respectively. The penetration of
these spin-singlet superconducting correlations are strongly supressed
by the ferromagnetic exchange field and are only short-range. In
addition, the exchange field also generates $S_{z} = 0$ (unequal-spin)
components of spin-triplet correlations, which also oscillate and
decay. Similar to the spin-singlet correlations these are short-range
as well.

However, it has been suggested theoretically by Bergeret \textit{et
  al.}~\cite{Bergeret_PRL86_4096}, that due to spin-flip processes at
the interface equal-spin spin-triplet Cooper pairs can form which
should be unaffected by the ferromagnetic exchange field thereby
allowing much larger penetration depths. This phenomenon is called the
long-range proximity effect and has triggered a lot of experimental
and theoretical work. The proximity effect in
superconductor-ferromagnet heterostructures is reviewed by
Buzdin~\cite{Buzdin_RMP77_935}, whereas Bergeret \textit{et
  al.}~\cite{Bergeret_RMP77_1321} review the physics behind this type
of ``odd-triplet'' superconductivity. The symmetry relations between
the different pairing correlations within the
superconductor-ferromagnet heterostructures is discussed in detail by
Eschrig~\textit{et al.}~\cite{Eschrig_JLowTempPhys147_457}.

From the experimental side several multilayer setups have been
suggested to observe spin-triplet proximity effect, but so far there
exist only indirect proofs of the generation of spin-triplet
supercurrents through the observation of supercurrents in Josephson
junctions~\cite{Klose_PRL108_127002,Gingrich_PRB86_224506} or
spin-valves~\cite{Leksin_PRL109_057005}.

Typical examples of multilayer systems to generate spin-triplet Cooper
pairs experimentally involve noncollinear magnetisations within the
different ferromagnetic
layers~\cite{Klose_PRL108_127002,Leksin_PRL109_057005,Gingrich_PRB86_224506},
or helical (or conical) magnets in the multilayer
setup~\cite{Halasz_PRB79_224505,Robinson_Science329_59,Halasz_PRB84_024517}. This
is accompanied by respective theoretical investigations of those
sytems containing noncollinear
magnetisations~\cite{Halterman_PRL99_127002,Halterman_PRB77_174511}
and helical (or conical) magnetic material in the multilayer
setup~\cite{Wu_PRB86_184517}. Additionally, also the effects of
Bloch~\cite{Alidoust_PRB81_014512} or
N\'{e}el~\cite{Fominov_PRB75_104509} domain walls, spin-orbit
coupling~\cite{Lv_EurPhysJB83_493} or specific interface
potentials~\cite{Bozovic_EPL70_513,Terrade_PRB88_054516} at the SC/FM
interface on the generation of spin-triplet Cooper pairs have been
investigated theoretically. Another route towards generating
spin-triplet Cooper pairs leads to the inclusion of half-metallic
ferromagnets such as CrO$_2$ into the
heterostructures~\cite{Keizer_Nature439_825,Eschrig_NaturePhys4_138,Anwar_PRB82_100501}
which would pave the way for a marriage between supercurrents and
spintronics applications~\cite{Eschrig_PhysToday64_43}.

The aim of the paper is as follows. A heterostructural setup similar
to those used in the experiments of Robinson \textit{et
  al.}~\cite{Robinson_Science329_59} consisting of superconductor,
conical magnet (CM) and ferromagnet will be investigated using
self-consistent solutions to the spin-dependent microscopic
Bogoliubov$-$de Gennes (BdG) equations~\cite{Annett_book}. One focus
of the work lies on the influence of the conical magnet's opening and
turning angles on the induced spin-triplet pairing correlations for
which a detailed symmetry analysis will be provided. Secondly, we
focus on the influence of conical magnetic layer thickness on the
spin-triplet correlations. It will be shown that a minimum number of
conical magnetic layers are necessary to efficiently generate
equal-spin spin-triplet correlations, in agreement with experimental
observations.

The paper is organised as follows. \Sref{Sec2} starts with a
description of the theoretical method, namely the self-consistent
solution of the spin-dependent BdG equations. This is followed by a
detailed description of the multilayer structure used in the
calculations, a setup for the conical magnetic structure, and finally
the spin-dependent pairing correlations. Results are presented in
\sref{Sec3}, where first a symmetry analysis of different conical
magnet orientations is presented followed by the conical magnet's and
ferromagnet's thickness dependence of the spin-dependent triplet
pairing correlations. A concluding summary and an outlook will be
given in \sref{SummaryAndOutlook}.

\section{Theoretical background and computational details}
\label{Sec2}

\subsection{Bogoliubov-de Gennes equations and tight-binding Hamiltonian}
\label{Sec2_1}

All our calculations are based on self-consistent solutions of the
microscopic BdG equations which for the spin-dependent case
read~\cite{Annett_book,Powell_JPhysA36_9289,KettersonSong_Superconductivity}
\begin{eqnarray}
\label{EqBdGGeneral}
\fl \left(
  \begin{array}{cccc}
    {\cal H}_{0} - h_{z} & -h_{x} + i h_{y} & \Delta_{\uparrow
      \uparrow} & \Delta_{\uparrow \downarrow} \\ - h_{x} - i h_{y} &
    {\cal H}_{0} + h_{z} & \Delta_{\downarrow \uparrow} &
    \Delta_{\downarrow \downarrow} \\ \Delta_{\uparrow \uparrow}^{*} &
    \Delta_{\downarrow \uparrow}^{*} & -{\cal H}_{0} + h_{z} & h_{x} +
    i h_{y} \\ \Delta_{\uparrow \downarrow}^{*} & \Delta_{\downarrow
      \downarrow}^{*} & h_{x} - i h_{y} & -{\cal H}_{0} - h_{z}
  \end{array}
  \right) \left(
  \begin{array}{c}
    u_{n\uparrow} \\ u_{n\downarrow} \\ v_{n\uparrow}
    \\ v_{n\downarrow}
  \end{array}
  \right) =\varepsilon_{n} \left(
  \begin{array}{c}
    u_{n\uparrow} \\ u_{n\downarrow} \\ v_{n\uparrow}
    \\ v_{n\downarrow}
  \end{array}
  \right) \,,
\end{eqnarray}
where $\varepsilon_{n}$ denote the eigenvalues of the matrix equation,
and $u_{n\sigma}$ and $v_{n\sigma}$ are the quasiparticle and
quasihole amplitudes for spin $\sigma$, respectively. ${\cal H}_{0}$
is the tight-binding Hamiltonian, which for a system of
two-dimensional layers can be written as
\begin{equation}
\label{EqTBHamiltonian}
  {\cal H}({\bf k}) = -t \sum_{n{\bf k}}{ \left( c_{n{\bf
        k}}^{\dagger}c_{n+1{\bf k}} + c_{n+1{\bf k}}^{\dagger}c_{n{\bf
        k}} \right) } + \sum_{n{\bf k}}{ \left( \varepsilon_{n{\bf k}}
    - \mu \right) c_{n{\bf k}}^{\dagger} c_{n{\bf k}} } \,,
\end{equation}
with $t$ being the next-nearest neighbour hopping parameter setting
the energy scale with $t = 1$, and $\mu = 0$ being the chemical
potential (Fermi energy) set to half-filling. $c_{n{\bf k}}^{\dagger}$
and $c_{n{\bf k}}$ are electronic creation and destruction operators
at layer index $n$ with momentum $\hbar {\bf k}$ within the layers,
respectively. Since the main focus of the present work lies on the
presence of an interface within the multilayer or heterostructure, the
only valid ${\bf k}$ values in \eref{EqTBHamiltonian} are to be found
within the interface plane. For each of these ${\bf k}$ values the BdG
equations \eref{EqBdGGeneral} leads to a one-dimensional inhomogeneous
problem in the layer index $n$~\cite{Covaci_PRB73_014503}. For the
sake of simplicity and since this would only lead to a parametrical
dependence of the Hamiltonian on a discretised ${\bf k}$ mesh in the
present work we neglect this ${\bf k}$
dependence. \Eref{EqTBHamiltonian} then simplifies to
\begin{equation}
\label{EqTBHamiltonianLinear}
  {\cal H}_{0} = -t \sum_{n}{ \left( c_{n}^{\dagger}c_{n+1} +
    c_{n+1}^{\dagger}c_{n} \right) } + \sum_{n}{ \left(
    \varepsilon_{n} - \mu \right) c_{n}^{\dagger} c_{n} } \,.
\end{equation}
The implications of this simplification will be taken into account
when discussing the obtained results in \sref{Sec3}.

The pairing matrix can be rewritten according to the
Balian$-$Werthamer
transformation~\cite{Balian_PhysRev131_1553,Sigrist_RMP63_239}
utilising the Pauli matrices $\sigma$
\begin{equation}
\label{EqBalianWerthamer}
\left( \begin{array}{cc} \Delta_{\uparrow \uparrow} & \Delta_{\uparrow
    \downarrow} \\ \Delta_{\downarrow \uparrow} & \Delta_{\downarrow
    \downarrow} \end{array} \right) = \left( \Delta + {\bf \sigma}{\bf
  d}\right) i \sigma_{2} = \left( \begin{array}{cc} -d_{x} + i d_{y} &
  \Delta + d_{z} \\ - \Delta + d_{z} & d_{x} + i d_{y} \end{array}
\right) \,,
\end{equation}
which effectively describes the superconducting order parameter
comprising of a singlet (scalar) part $\Delta$ and a triplet (vector)
part ${\bf d}$, respectively. In the present work $\Delta$ is
restricted to the $s$-wave singlet pairing potential in the
superconductor sides of the heterostructure, to be determined
self-consistently from the condition
\begin{equation}
\Delta({\bf r}) = \frac{g({\bf r})}{2} \sum_{n}{ \bigl(
  u_{n\uparrow}({\bf r})v_{n\downarrow}^{*}({\bf r})
  [1-f(\varepsilon_{n})] + u_{n\downarrow}({\bf
    r})v_{n\uparrow}^{*}({\bf r}) f(\varepsilon_{n}) \bigr)} \,,
\end{equation}
where the summation is performed over the positive eigenvalues
$\varepsilon_{n}$. $f(\varepsilon_{n})$ is the Fermi distribution
function and $g({\bf r})$ the effective superconducting coupling set
to $1$ in our calculations. It is assumed to be constant within the
superconductor and to vanish elsewhere.

Finally, $h_{x}$, $h_{y}$, and $h_{z}$ generally describe the vector
components of a noncollinear exchange field to be added to the
tight-binding Hamiltonian in the form ${\bf{h}} {\bf{\sigma}}$, with
the vector components of ${\bf{\sigma}}$ being the Pauli matrices,
respectively. In \sref{Sec2_2} ${\bf{h}}$ will be defined to describe
the conical magnetic structure within the multilayer setup.

\subsection{Multilayer structural setup}
\label{Sec2_2}

\begin{figure}
  \centering
  \includegraphics[width=0.75\columnwidth,clip]{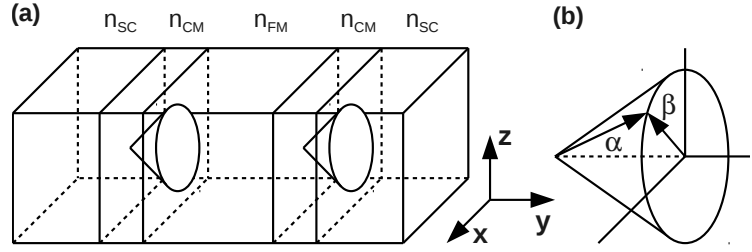}
  \fl \caption{\label{Fig1} (a) Multilayer structural setup consisting
    of a spin-singlet $s$-wave superconductor ($n_{\rm SC}$ layers), a
    conical magnet ($n_{\rm CM}$ layers), a ferromagnetic metal
    ($n_{\rm FM}$ layers), and a conical magnet and superconductor of
    the same thickness to the right. (b) Opening angle $\alpha$ and
    turning angle $\beta$ of the conical magnet. From
    \eref{EqHelicalMagnet} it follows that $\alpha$ is measured from
    the positive $y$ axis towards the positive $z$ axis, whereas
    $\beta$ is measured from the positive $z$ axis towards the
    positive $x$ axis.}
\end{figure}

The multilayer setup used in the present work is schematically shown
in \fref{Fig1}(a).
It consists of a spin-singlet $s$-wave superconductor of $n_{\rm
  SC}=250$ layers, a conical magnet of $n_{\rm CM}=1 \cdots 25$
layers, a ferromagnetic metal of up to $n_{\rm FM}=500$ layers,
followed by the same number of layers of conical magnet $n_{\rm CM}$
and spin-singlet $s$-wave superconductor $n_{\rm SC}$ to the right,
respectively. The description of the conical magnet is chosen
according to Wu \textit{et al.}~\cite{Wu_PRB86_184517}
\begin{equation}
\label{EqHelicalMagnet}
  {\bf h} = h_{0} \left\{ \cos \alpha {\bf y} + \sin \alpha \left[
    \sin \left( \frac{\beta y}{a} \right) {\bf x} + \cos \left(
    \frac{\beta y}{a} \right) {\bf z} \right] \right\} \,,
\end{equation}
with $h_{0}$ being the strength of the conical magnet's exchange field
and $a$ being the lattice constant (set to unity $a=1$). As can be
seen from \eref{EqHelicalMagnet} and \fref{Fig1}(b), the opening angle
$\alpha$ is measured from the positive $y$ axis towards the positive
$z$ axis, whereas the turning angle $\beta$ is measured from the
positive $z$ axis towards the positive $x$ axis. Here these angles
have been kept fixed to the values $\alpha = 80\,^{\circ}$ and $\beta
= 30\,^{\circ}$ to represent the conical magnet Holmium, a transition
metal routinely used in similar experimental
investigations~\cite{Robinson_Science329_59}. Since the experimental
geometry of how the conical structure is oriented with respect to the
ferromagnetic region is unknown~\cite{Robinson_PrivComm} our first set
of calculations will examine the effects of different orientations and
turning angle directions of the conical structure with respect to the
two different ferromagnetic interfaces in \sref{Sec3_1}.

\subsection{(Triplet) Pairing correlations}
\label{Sec2_3}

The general expression for the on-site superconducting pairing
correlation of spins $\alpha$ and $\beta$ for times $t = \tau$ and $t'
= 0$ reads
\begin{equation}
  f_{\alpha \beta}({\bf r}, \tau, 0) = \frac{1}{2}\bigl<
  \hat{\Psi}_{\alpha}({\bf r},\tau) \hat{\Psi}_{\beta}({\bf r},0)
  \bigr> \,.
\end{equation}
Therein, $\hat{\Psi}_{\sigma}({\bf r},\tau)$ denotes the many-body
field operator for spin $\sigma$ at time $\tau$, and the
time-dependence is governed by the Heisenberg equation of
motion. Notice that this pairing correlation is local in space and so
the triplet contributions vanish automatically in the case $\tau = 0$
according to the Pauli
principle~\cite{Halterman_PRL99_127002}. Therefore, such a pairing
field is only non-zero at finite times $\tau$, an example of
odd-frequency triplet pairing~\cite{Bergeret_RMP77_1321}. Substituting
the field operators valid for our setup and phase convention the
spin-dependent triplet pairing correlations read
\begin{equation}
\label{EqTripletPairingCorrelations}
  \eqalign { f_{\uparrow \downarrow}(y,\tau) + f_{\downarrow
      \uparrow}(y,\tau) = \frac{1}{2} \sum_{n}{\bigl( u_{n\uparrow}(y)
      v_{n\downarrow}^{*}(y) + u_{n\downarrow}(y) v_{n\uparrow}^{*}(y)
      \bigr) \zeta_{n}(\tau)} \cr f_{\uparrow \uparrow}(y,\tau) =
    \frac{1}{2} \sum_{n} {\bigl( u_{n\uparrow}(y) v_{n\uparrow}^{*}(y)
      \bigr) \zeta_{n}(\tau)} \cr f_{\downarrow \downarrow}(y,\tau) =
    \frac{1}{2} \sum_{n} {\bigl( u_{n\downarrow}(y)
      v_{n\downarrow}^{*}(y) \bigr) \zeta_{n}(\tau)} }
\end{equation}
depending on the time parameter $\tau$ and with $\zeta_{n}(\tau)$
given by
\begin{equation}
\label{EqTau}
  \zeta_{n}(\tau) = \cos (\varepsilon_{n} \tau) - i \sin
  (\varepsilon_{n} \tau) \rm{tanh} \left( \frac{\varepsilon_{n}}{2T}
  \right) \,.
\end{equation}
These triplet pairing correlations correspond to $S_{z} = 0$
$(f_{\uparrow \downarrow} + f_{\downarrow \uparrow})$, +1
$(f_{\uparrow \uparrow})$, and -1 $(f_{\downarrow \downarrow})$,
respectively.

\section{Results and discussion}
\label{Sec3}

\subsection{Influence of conical magnet}
\label{Sec3_1}

As soon as a conical magnetic structure is included in the multilayer
setup there are several ways to orient the magnetic moments with
respect to the direction perpendicular to the interface layer (being
the $y$-axis in our multilayer setup). As mentioned earlier, the
opening angle $\alpha = 80\,^{\circ}$ and turning angle $\beta =
30\,^{\circ}$ of the conical magnet are chosen to represent the
magnetic structure of Holmium, routinely used in experimentally
available multilayer structures~\cite{Robinson_PrivComm}. Experimental
evidence shows that the magnetic coupling at the CM/FM interface is
most likely antiferromagnetic. Looking for the moment at the right
FM/CM interface (\fref{Fig1}(a)) and assuming the ferromagnetic
moments to orient along the $+z$ axis, the conical magnetic moment
closest to the interface can have two different antiferromagnetic-like
orientations, namely pointing slightly towards the FM side
($\alpha_{\rm R} = 260\,^{\circ}$ case, with the cone opening into the
FM layer) or slightly towards the CM side of the interface
($\alpha_{\rm R} = 280\,^{\circ}$ case, with the cone opening away
from the FM layer). These angles are reversed at the left CM/FM
interface, respectively.

Furthermore, the handedness of the respective cone is determined not
only by the turning angle $\beta_{\rm R}$ ($30\,^{\circ}$ and
$-30\,^{\circ}$) but also influenced by the respective opening
angle. Looking again at the right FM/CM interface and an opening angle
$\alpha_{\rm R} = 280\,^{\circ}$ (cone opening away from the FM layer)
a turning angle $\beta_{\rm R} = 30\,^{\circ}$ ($\beta_{\rm R} =
-30\,^{\circ}$) describes a clockwise (counterclockwise) rotation of
the conical magnetic structure. If, however, the opening angle amounts
to $\alpha_{\rm R} = 260\,^{\circ}$ (cone opening into the FM layer),
a turning angle $\beta_{\rm R} = 30\,^{\circ}$ ($-30\,^{\circ}$) again
describes a clockwise (counterclockwise) rotation but now viewed along
the $-y$ direction.

\begin{figure}
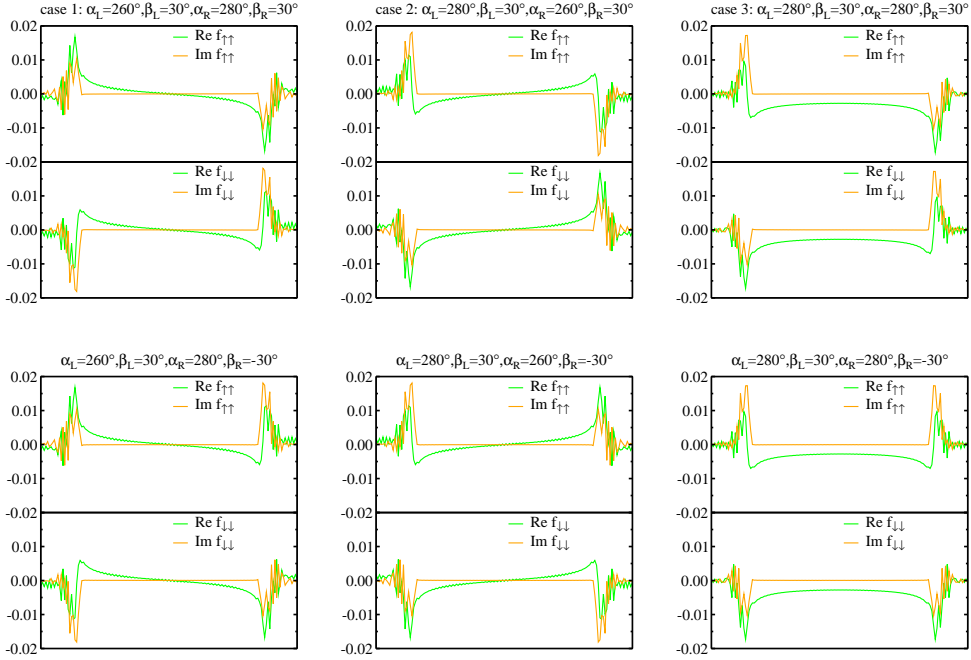

  \begin{center}
    \subfigure{
      \includegraphics[width=0.3\textwidth,clip]{./figure2a.eps}
      \label{Fig2:subfig1}}
    \hfill \subfigure{
      \includegraphics[width=0.3\textwidth,clip]{./figure2b.eps}
      \label{Fig2:subfig2}}
    \hfill \subfigure{
      \includegraphics[width=0.3\textwidth,clip]{./figure2c.eps}
      \label{Fig2:subfig3}}
  \end{center}
  \begin{center}
    \subfigure{
      \includegraphics[width=0.3\textwidth,clip]{./figure2d.eps}
      \label{Fig2:subfig4}}
    \hfill \subfigure{
      \includegraphics[width=0.3\textwidth,clip]{./figure2e.eps}
      \label{Fig2:subfig5}}
    \hfill \subfigure{
      \includegraphics[width=0.3\textwidth,clip]{./figure2f.eps}
      \label{Fig2:subfig6}}
  \end{center}
  \caption{\label{Fig2} Influence of different opening angles of the
    left ($\alpha_{\rm L}$) and right ($\alpha_{\rm R}$) CM structure
    adjacent to the middle FM layer on the real (green) and imaginary
    part (orange) of the equal-spin spin-triplet pairing correlations
    $f_{\uparrow \uparrow}$ and $f_{\downarrow \downarrow}$. Upper and
    lower panels depict the influence of equal ($\beta_{\rm L} =
    \beta_{\rm R} = 30\,^{\circ}$) and opposite ($\beta_{\rm L} =
    -\beta_{\rm R} = 30\,^{\circ}$) handedness of the two CM
    structures. The definition of $\alpha$ and $\beta$ is given
    according to \eref{EqHelicalMagnet} and \fref{Fig1}(b).}
\end{figure}

The influence of different opening angles of the conical magnetic
layers on both sides of the ferromagnetic region on the triplet
pairing correlations $f_{\uparrow \uparrow}$ and $f_{\downarrow
  \downarrow}$ are shown in the left, middle, and right panels of
\fref{Fig2}, whereas upper and lower panels depict the influence of
equal and opposite handedness of the two conical magnetic structures,
respectively.
The relation between the triplet pairing correlations $f_{\uparrow
  \uparrow}$ and $f_{\downarrow \downarrow}$ depending on different
choices are given in \tref{TabSCSpiralTest}.
\begin{table}
  \caption{\label{TabSCSpiralTest} Superconductor-conical magnet
    interface symmetry properties depending on the two conical
    magnet's angles. Given is the relation of the left-side triplet
    pairing correlations $f_{\uparrow \uparrow}^{\rm L}$ and
    $f_{\downarrow \downarrow}^{\rm L}$ (first column) to the
    corresponding right-side triplet pairing correlation $f_{\uparrow
      \uparrow}^{\rm R}$ and $f_{\downarrow \downarrow}^{\rm R}$
    depending on the opening angles $\alpha_{\rm L}$ and $\alpha_{\rm
      R}$, and the turning angles $\beta_{\rm L}$ and $\beta_{\rm R}$,
    respectively. As can also be seen from \fref{Fig2} the given
    dependencies apply equally for the real and imaginary part.}
  \begin{indented}
  \item[] \begin{tabular}{cccc} \br

    & case 1: & case 2: & case 3: \\
    
    & $\alpha_{\rm L}=260^{\circ}$, $\alpha_{\rm R}=280^{\circ}$ &
    $\alpha_{\rm L}=280^{\circ}$, $\alpha_{\rm R}=260^{\circ}$ &
    $\alpha_{\rm L}=280^{\circ}$, $\alpha_{\rm R}=280^{\circ}$ \\
      
    \ns
    
    & \crule{1} & \crule{1} & \crule{1} \\
    
    & $\beta_{\rm L}=30^{\circ}$, $\beta_{\rm R}=30^{\circ}$ &
    $\beta_{\rm L}=30^{\circ}$, $\beta_{\rm R}=30^{\circ}$ &
    $\beta_{\rm L}=30^{\circ}$, $\beta_{\rm R}=30^{\circ}$ \\
    
    \mr
    
    $f_{\uparrow \uparrow}^{\rm L}$ & \centering $-f_{\uparrow
      \uparrow}^{\rm R}$ & $-f_{\uparrow \uparrow}^{\rm R}$ &
    $f_{\downarrow \downarrow}^{\rm R}$ \\
      
    \ms
    
    $f_{\downarrow \downarrow}^{\rm L}$ & $-f_{\downarrow
      \downarrow}^{\rm R}$ & $-f_{\downarrow \downarrow}^{\rm R}$ &
    $f_{\uparrow \uparrow}^{\rm R}$ \\
    
    \mr 
    
    & $\beta_{\rm L} = 30^{\circ}$, $\beta_{\rm R} = -30^{\circ}$ &
    $\beta_{\rm L} = 30^{\circ}$, $\beta_{\rm R} = -30^{\circ}$ &
    $\beta_{\rm L} = 30^{\circ}$, $\beta_{\rm R} = -30^{\circ}$ \\
    
    \mr
    
    $f_{\uparrow \uparrow}^{\rm L}$ & $-f_{\downarrow \downarrow}^{\rm
      R}$ & $-f_{\downarrow \downarrow}^{\rm R}$ & $f_{\uparrow
      \uparrow}^{\rm R}$ \\
    
    \ms
      
    $f_{\downarrow \downarrow}^{\rm L}$ & $-f_{\uparrow \uparrow}^{\rm
      R}$ & $-f_{\uparrow \uparrow}^{\rm R}$ & $f_{\downarrow
      \downarrow}^{\rm R}$ \\
    
    \br
  \end{tabular}
  \end{indented}
\end{table}
Looking for the moment only at case 1 in \fref{Fig2} and
\tref{TabSCSpiralTest} ($\alpha_{\rm L} = 260\,^{\circ}$, $\alpha_{\rm
  R} = 280\,^{\circ}$) with $\beta_{\rm L} = \beta_{\rm R} =
30\,^{\circ}$. From the symmetry discussion of the two cones it's
apparent that the conical magnetic structure left of the FM interface
is opening away from the interface towards the $-y$ direction with a
clockwise rotation. The same holds for the right side of the
interface; the cone is opening away from the interface towards the $y$
direction with a clockwise rotation. For this setup both conical
magnetic structures seem to be identical; they both open away from the
FM interface into the CM layers with a clockwise rotation of the
conical magnetisation. But the results for this setup show a sign
change between the left and right side CM/FM interfaces (\fref{Fig2}
and \tref{TabSCSpiralTest}) with $f_{\uparrow \uparrow}^{\rm L} = -
f_{\uparrow \uparrow}^{\rm R}$ and $f_{\downarrow \downarrow}^{\rm L}
= - f_{\downarrow \downarrow}^{\rm R}$. At first this looks like a
discrepancy, but in fact this stems from the underlying symmetry of
the ${\bf d}({\bf r})$ vector describing the triplet pairing
correlations which will be discussed now. Although in case 1 and
$\beta_{\rm L} = \beta_{\rm R} = 30\,^{\circ}$ both conical magnetic
structures seem to be identical, in fact they can be transformed into
one another by a $C_{2}$ rotation about the $z$ axis located in the
middle of the FM layers. According to Tinkham
\cite{Tinkham_GroupTheory} the transformation of an arbitrary vector
${\bf r}$ under a symmetry operation described by a transformation
matrix $R(u)$ reads
\begin{equation}
r_{i}^{'} = \sum_{j}{R(u)_{j}r_{j}} \,.
\end{equation}
Applying the transformation matrix for a $C_{2}$ rotation given by
\begin{eqnarray}
  R(C_{2}) = \left( \begin{array}{ccc} -1 & 0 & 0 \\ 0 & -1 & 0 \\ 0 &
    0 & 1 \end{array} \right) \,.
\end{eqnarray}
to the superconducting order parameter written in the
Balian$-$Werthamer way as of \eref{EqBalianWerthamer} yields
\begin{equation}
\eqalign { R(C_{2}) \left( \begin{array}{cc} \Delta_{\uparrow
      \uparrow} & \Delta_{\uparrow \downarrow} \\ \Delta_{\downarrow
      \uparrow} & \Delta_{\downarrow \downarrow} \end{array} \right) =
  R(C_{2}) \left( \begin{array}{cc} -d_{x} + i d_{y} & \Delta + d_{z}
    \\ \Delta + d_{z} & d_{x} + i d_{y} \end{array} \right) \cr =
  \left( \begin{array}{cc} d_{x} - i d_{y} & \Delta + d_{z} \\ \Delta
    + d_{z} & - d_{x} - i d_{y} \end{array} \right) =
  \left( \begin{array}{cc} -\Delta_{\uparrow \uparrow} &
    \Delta_{\uparrow \downarrow} \\ \Delta_{\downarrow \uparrow} & -
    \Delta_{\downarrow \downarrow} \end{array} \right) } \,,
\end{equation}
and expresses exactly what is displayed in \fref{Fig3} and given in
\tref{TabSCSpiralTest}, namely $f_{\uparrow \uparrow}^{\rm L} = -
f_{\uparrow \uparrow}^{\rm R}$ and $f_{\downarrow \downarrow}^{\rm L}
= - f_{\downarrow \downarrow}^{\rm R}$, respectively.

Looking now at case 1 but for the two cones having different
handednesses (lower left panels of \fref{Fig2} and
\tref{TabSCSpiralTest}) one notices a mixture between $\uparrow
\uparrow$ and $\downarrow \downarrow$ contributions, namely
$f_{\uparrow \uparrow}^{\rm L} = - f_{\downarrow \downarrow}^{\rm R}$
and $f_{\downarrow \downarrow}^{\rm L} = - f_{\uparrow \uparrow}^{\rm
  R}$, respectively. In this case the transformation between the left
and right conical magnetic structure is realised by a $\sigma_{xz}$
mirror plane again located in the middle of the FM layers with the
respective transformation matrix
\begin{eqnarray}
  R(\sigma_{xz}) = \left( \begin{array}{ccc} 1 & 0 & 0 \\ 0 & -1 & 0
    \\ 0 & 0 & 1 \end{array} \right) \,.
\end{eqnarray}
Applying $R(\sigma_{xz})$ to the superconducting order parameter yields
\begin{equation}
R(\sigma_{xz}) \left( \begin{array}{cc} \Delta_{\uparrow \uparrow} &
  \Delta_{\uparrow \downarrow} \\ \Delta_{\downarrow \uparrow} &
  \Delta_{\downarrow \downarrow} \end{array} \right) =
\left( \begin{array}{cc} - \Delta_{\downarrow \downarrow} &
  \Delta_{\uparrow \downarrow} \\ \Delta_{\downarrow \uparrow} & -
  \Delta_{\uparrow \uparrow} \end{array} \right) \,,
\end{equation}
in agreement with results displayed in \fref{Fig2} and given in
\tref{TabSCSpiralTest}. Again, the sign changes and swapping of
contributions for $\uparrow \uparrow$ and $\downarrow \downarrow$
between the left and right conical magnetic structures reveal the
underlying symmetry properties of the ${\bf d}({\bf r})$ vector for
the chosen multilayer setup. Now similar arguments along those lines
explain the results for case 2 shown in the middle panels of
\fref{Fig3} and \tref{TabSCSpiralTest}, respectively. Summarising
this, opposite opening angles on both interfaces just give a sign
change, whereas a different handedness in addition mixes the $\uparrow
\uparrow$ and $\downarrow \downarrow$ contributions. Although there
are no symmetry arguments available for case 3 shown in the right
panels of \fref{Fig2} and \tref{TabSCSpiralTest} one can understand
the results on the basis of symmetry arguments provided from case 1
and 2 above. The most striking difference in case 3 is the nodeless
behaviour of the $\uparrow \uparrow$ and $\downarrow \downarrow$
contributions.

There is nearly no influence of different $\alpha$s and $\beta$s on
the $f_{\uparrow \downarrow}+f_{\downarrow \uparrow}$ spin-triplet
pairing correlation corresponding to the $S_{z} = 0$ case (as depicted
in the middle panel of \fref{Fig3}).

\subsection{Influence of time parameter $\tau$ on spin-triplet pairing correlations}
\label{Sec3_2}

\begin{figure}
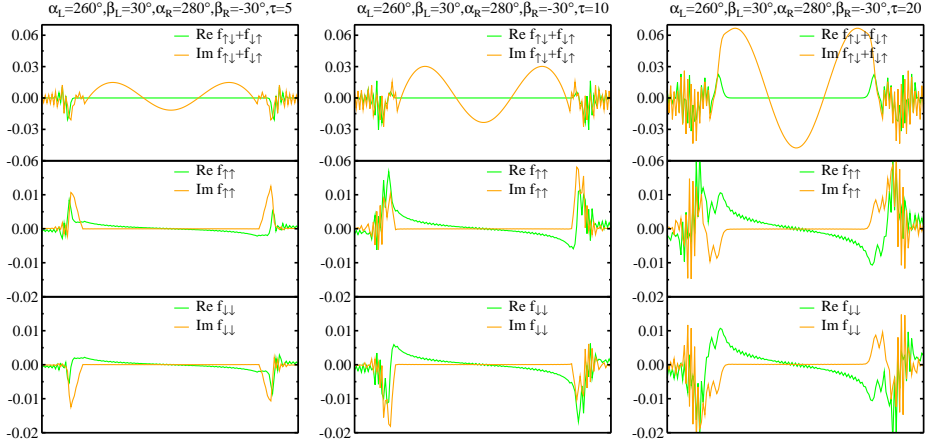

\centering \subfigure{
  \includegraphics[width=0.3\textwidth,clip]{./figure3a.eps}}
\subfigure{
  \includegraphics[width=0.3\textwidth,clip]{./figure3b.eps}}
\subfigure{
  \includegraphics[width=0.3\textwidth,clip]{./figure3c.eps}}
\caption{\label{Fig3} Influence of the time parameter $\tau$ entering
  \eref{EqTau} on the real (green) and imaginary part (orange) of the
  triplet pairing correlations $f_{\uparrow \downarrow} +
  f_{\downarrow \uparrow}$ (upper panels), $f_{\uparrow \uparrow}$
  (middle panels) and $f_{\downarrow \downarrow}$ (lower panels),
  respectively. The definition of $\alpha$ and $\beta$ is given
  according to \eref{EqHelicalMagnet} and \fref{Fig1}(b).}
\end{figure}

Chosing a specific fixed setup (case 1 as mentioned in \sref{Sec3_1})
this section deals with the influence of the time parameter $\tau$
entering the evaluations of the spin-dependent triplet pairing
correlations \eref{EqTripletPairingCorrelations} utilising
\eref{EqTau}. In addition to \fref{Fig2} the real and imaginary parts
of $f_{\uparrow \downarrow}+f_{\downarrow \uparrow}$, $f_{\uparrow
  \uparrow}$, and $f_{\downarrow \downarrow}$ are shown in \fref{Fig3}
for different times $\tau$ ranging from $5$ to $20$.
The upper panels display the unequal spin-triplet pairing amplitudes
$f_{\uparrow \downarrow}+f_{\downarrow \uparrow}$. They clearly
exhibit the oscillating behaviour associated with FFLO oscillations
inside the FM region of the multilayer, whereas these oscillations are
absent for the spin-equal triplet pairing amplitudes $f_{\uparrow
  \uparrow}$ (middle panels) and $f_{\downarrow \downarrow}$ (lower
panels) of \fref{Fig3}. Concentrating for the moment on the middle
panels of \fref{Fig3} for $\tau=10$ one immediately recognises a
change by a factor of $1/2$ (left panels) and $2$ (right panels) in
the spin triplet pairing correlations when comparing with the left and
right panels showing results obtained for times $\tau$ which are also
changed by a factor of $1/2$ and $2$, respectively. Recognising this
essentially linear dependence on the time factor $\tau$ in the present
regime $\tau \times \Delta << 1$ entering the calculation of the spin
triplet pairing correlations via \eref{EqTau}, further calculations
are restricted to a time parameter $\tau = 10$. A more detailed
investigation of the influence of $\tau$ on the spin triplet pairing
correlations will be part of a later work. But here we simply note
again that this triplet pairing correlation which is spatially local
but retarded in time, vanishes at $\tau = 0$, corresponding to the
``odd triplet'' pairing state derived by quasiclassical arguments by
Bergeret \textit{et al.}~\cite{Bergeret_RMP77_1321} and
Eschrig~\textit{et al.}~\cite{Eschrig_JLowTempPhys147_457}.

\subsection{Influence of ferromagnetic layer thickness $n_{\rm FM}$ on spin-triplet pairing correlations}
\label{Sec3_3}

\begin{figure}
\centering \subfigure{
  \includegraphics[height=0.175\textheight,clip]{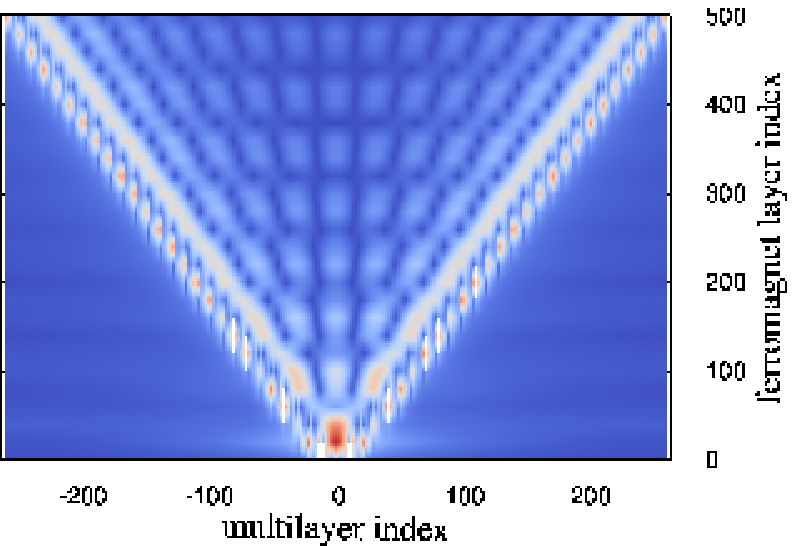}}
\subfigure{
  \includegraphics[height=0.175\textheight,clip]{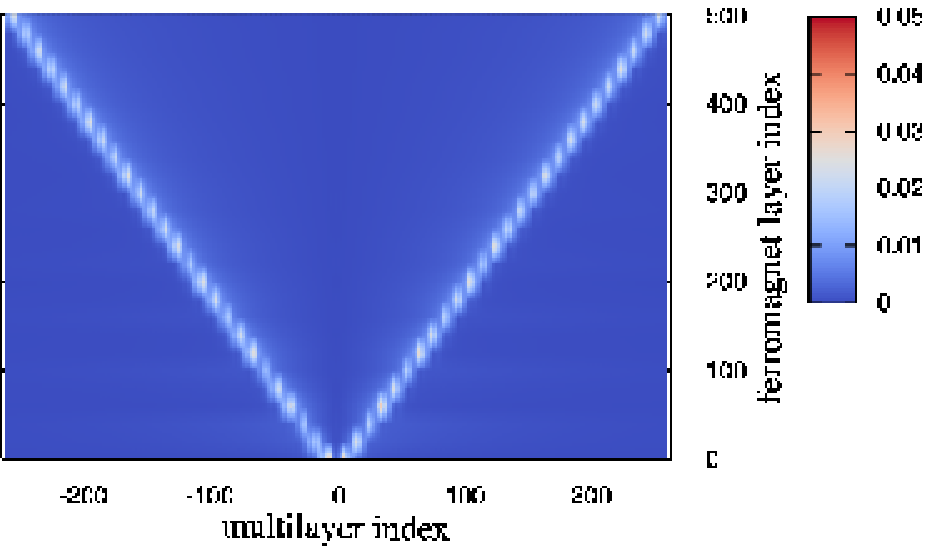}}
\\ \subfigure{
  \includegraphics[height=0.175\textheight,clip]{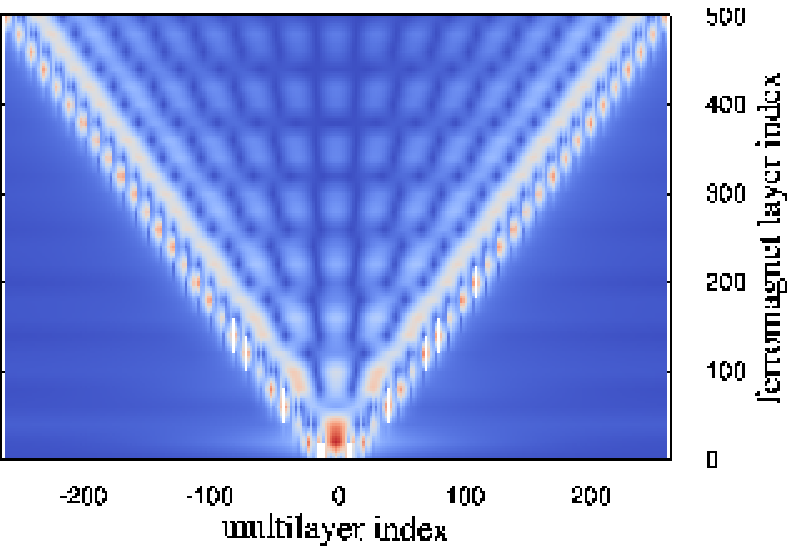}}
\subfigure{
  \includegraphics[height=0.175\textheight,clip]{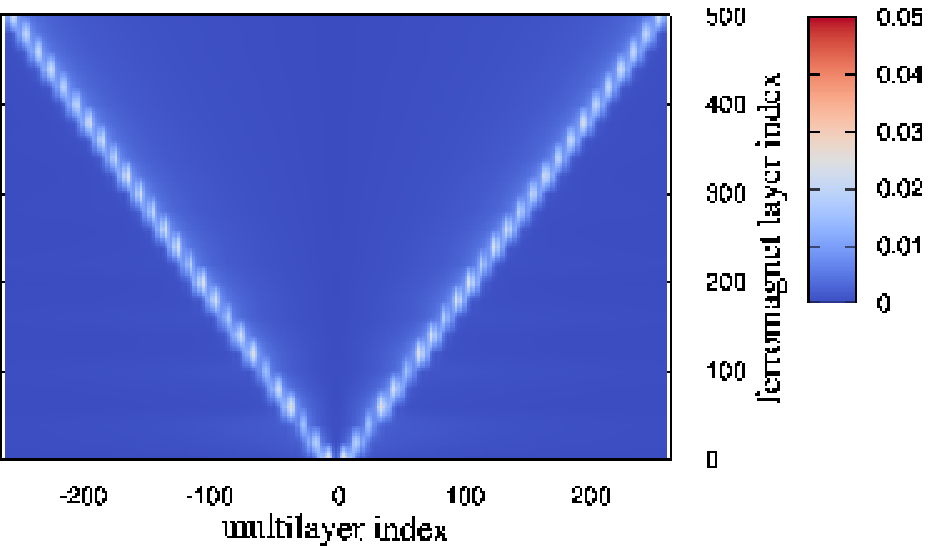}}
\\ \subfigure{
  \includegraphics[height=0.175\textheight,clip]{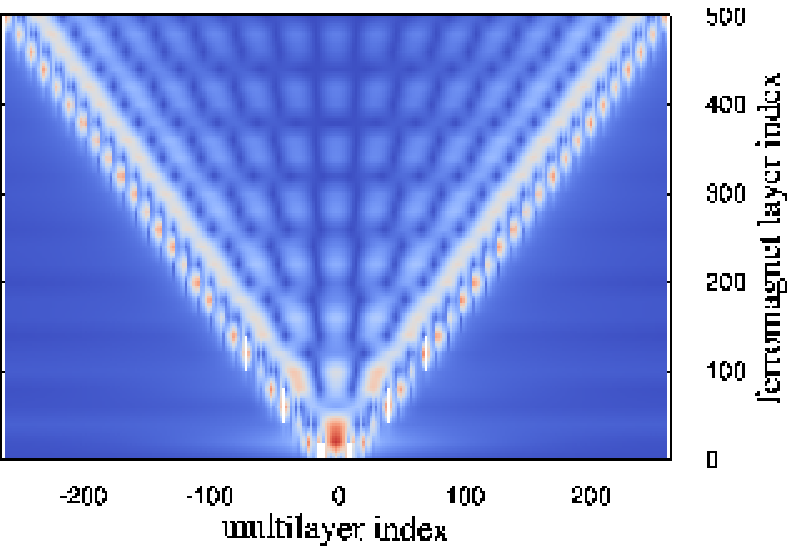}}
\subfigure{
  \includegraphics[height=0.175\textheight,clip]{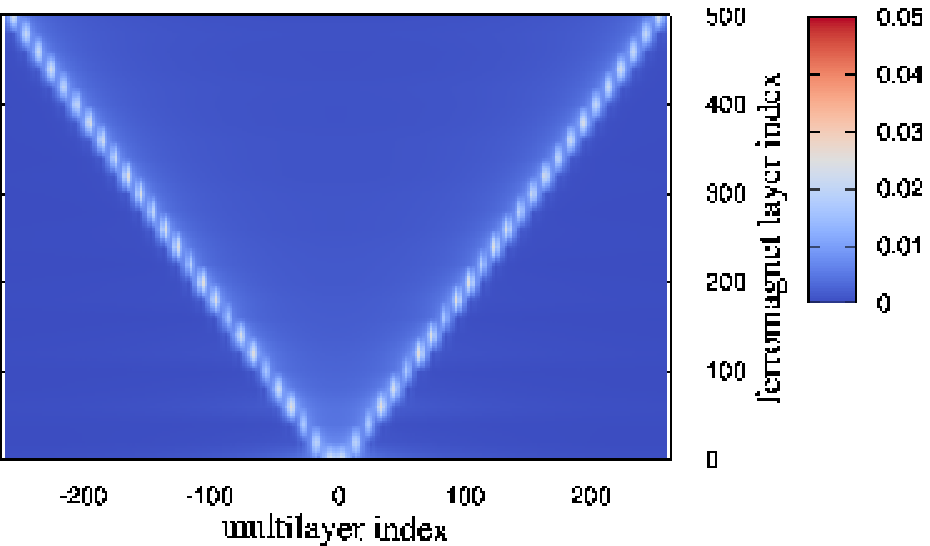}}
\caption{\label{Fig4} Influence of ferromagnetic layer thickness on
  the magnitude of spin-triplet pairing correlations. Shown are the
  $f_{\uparrow \downarrow}+f_{\downarrow \uparrow}$ (left panels) and
  $f_{\uparrow \uparrow}$ contributions (right panels) for multilayer
  setups case 1 (upper panels), case 2 (middle panels), and case 3
  (lower panels) as discussed in \sref{Sec3_1} and depicted in
  \fref{Fig2}, respectively.}
\end{figure}

Using the same geometries as above (case 1, case 2, and case 3 of
\sref{Sec3_1}) and one full conical magnetic structure on either side
of the ferromagnetic layer now the influence of the ferromagnet's
layer thickness on the spin-dependent triplet pairing correlations
shall be investigated. \Fref{Fig4} displays the magnitude of the
spin-triplet pairing correlations $f_{\uparrow
  \downarrow}+f_{\downarrow \uparrow}$ in the left and $f_{\uparrow
  \uparrow}$ contributions in the right panels for conical magnetic
orientations and multilayer setups according to case 1 (upper panels),
case 2 (middle panels), and case 3 (lower panels) as discussed in
\sref{Sec3_1} and depicted in \fref{Fig2}, respectively.
The $f_{\downarrow \downarrow}$ contributions are identical to the
$f_{\uparrow \uparrow}$ contributions and are not shown
here. Concentrating for a moment only on the left panels of
\fref{Fig4}, one notices the influence of increasing ferromagnetic
layer thickness as more and more oscillations are appearing in this
region. This $f_{\uparrow \downarrow}+f_{\downarrow \uparrow}$
contributions are unaffected by the specific multilayer setup. It
should be noted at this point that the very slowly decaying
$f_{\uparrow \downarrow}+f_{\downarrow \uparrow}$ spin-triplet pairing
correlations are an artefact of the simplified linear chain model used
in these calculations and decay much faster once the fully ${\bf k}$
dependent Hamiltonian \eref{EqTBHamiltonian} is used in the
calculations. However, the interest of the present work lies only on
the equal-spin spin-triplet correlations and how effectively they can
be generated at an interface containing a conical magnetic
structure. Looking now at the right panels of \fref{Fig4} one notices
the zeros along the line belonging to the middle multilayer index in
the upper two panels showing results for multilayer setups case 1 and
case 2. This is in line with \fref{Fig2} and shows that an increasing
ferromagnetic layer thickness does not give rise to more zeros in the
ferromagnetic region. The nonvanishing contributions present in
multilayer setup case 3 as shown in \fref{Fig2} (right panels) are
also present for increasing ferromagnetic thickness (lower right panel
of \fref{Fig4}). All in all there is no influence of the ferromagnetic
layer thickness on the behaviour of the equal-spin spin-triplet
correlations with respect to showing additional or less zeros. For the
next investigations the number of ferromagnetic layers will be fixed
to $n_{\rm FM} = 100$ layers, respectively.

\subsection{Influence of conical magnetic layer thickness $n_{\rm CM}$ on spin-triplet pairing correlations}
\label{Sec3_4}

\begin{figure}
\centering \subfigure{
  \includegraphics[width=0.485\textwidth,clip]{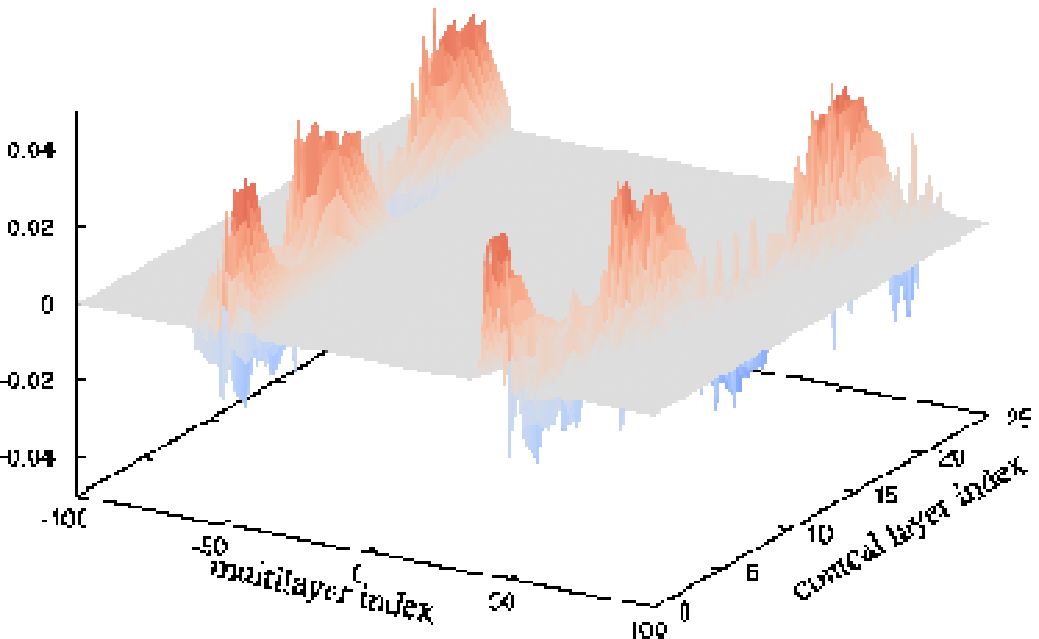}}
\subfigure{
  \includegraphics[width=0.485\textwidth,clip]{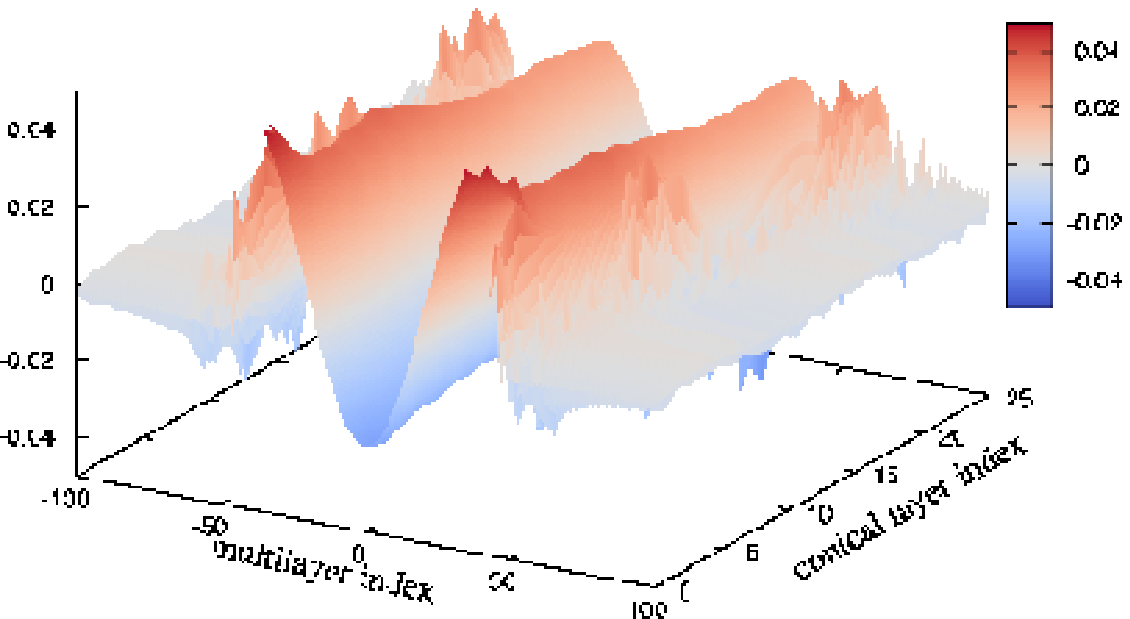}}
\\ \subfigure{
  \includegraphics[width=0.485\textwidth,clip]{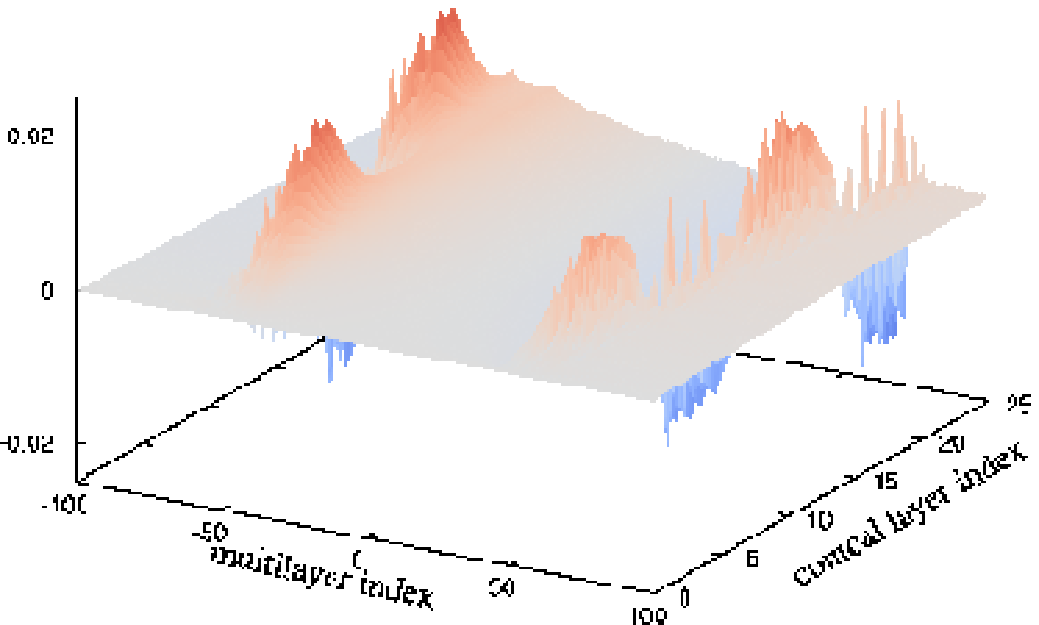}}
\subfigure{
  \includegraphics[width=0.485\textwidth,clip]{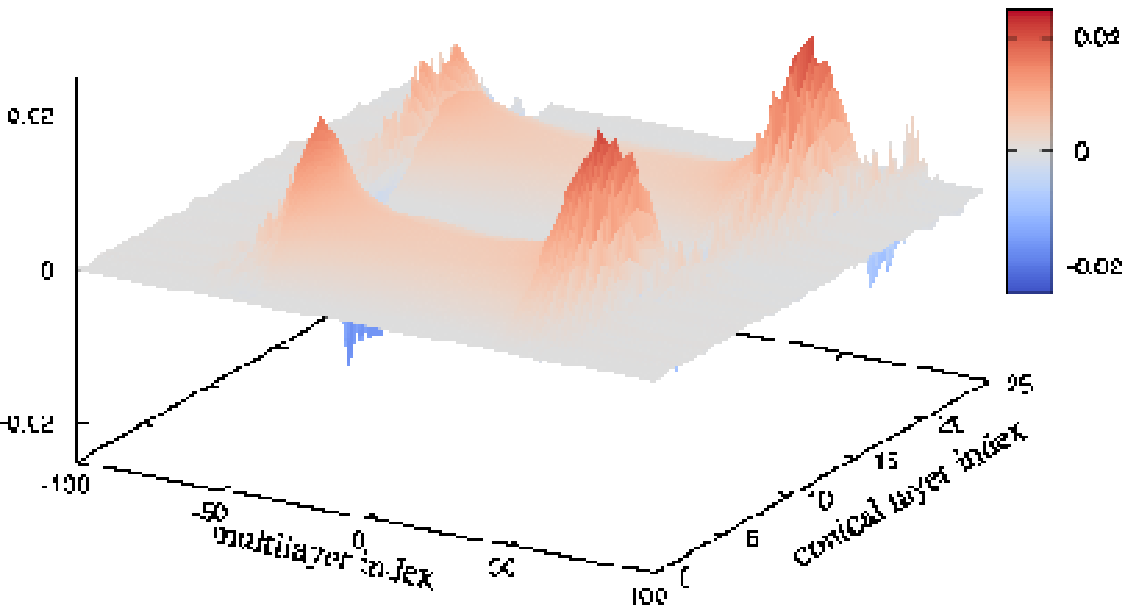}}
\\ \subfigure{
  \includegraphics[width=0.485\textwidth,clip]{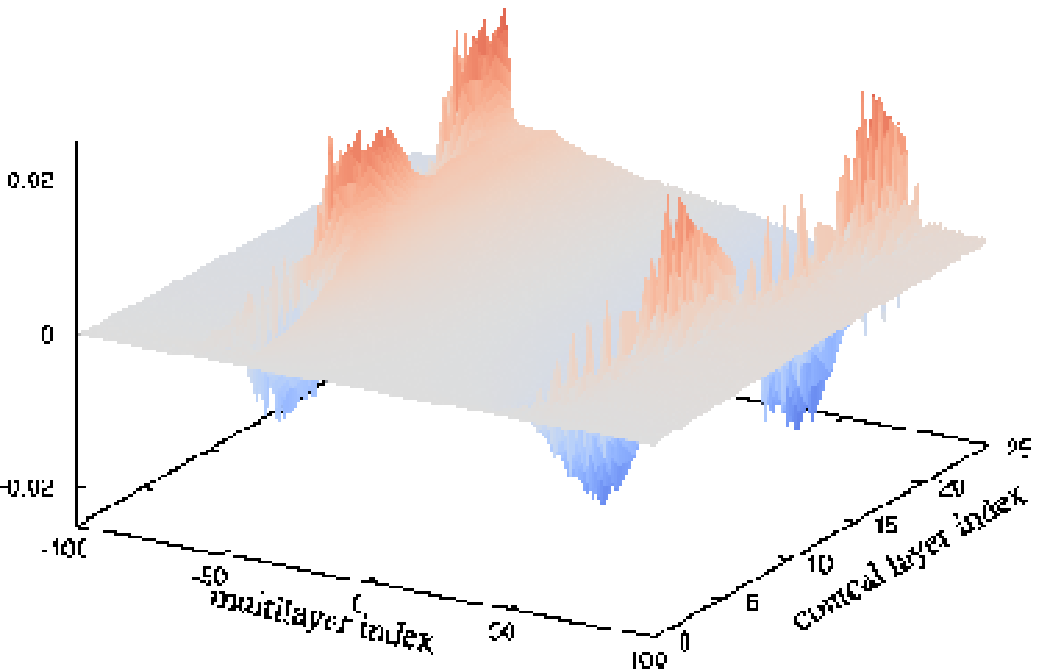}}
\subfigure{
  \includegraphics[width=0.485\textwidth,clip]{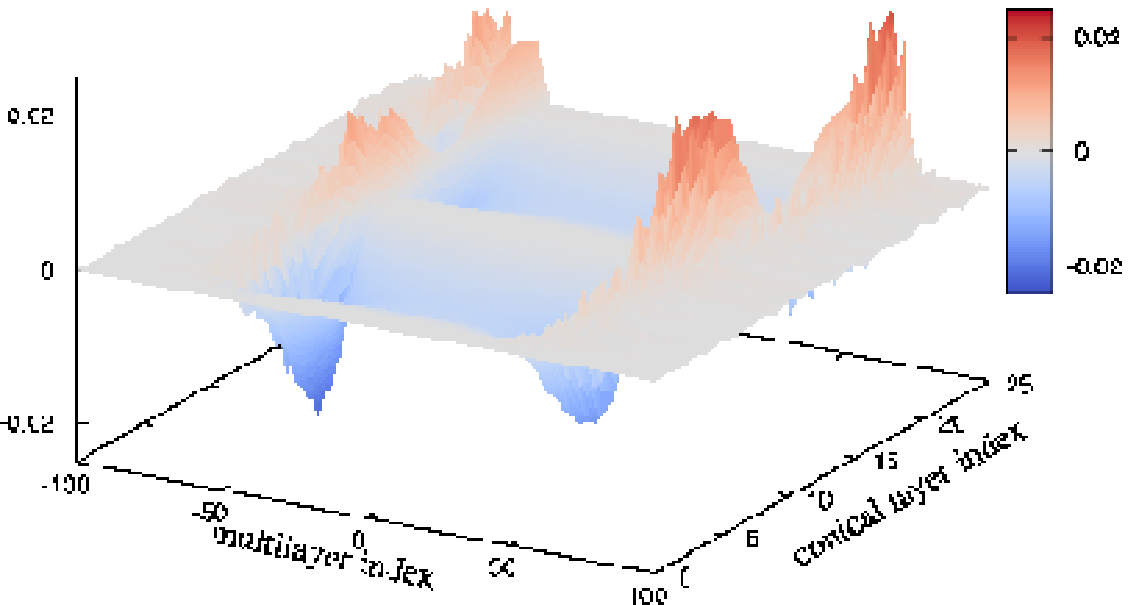}}
\caption{\label{Fig5} Influence of increasing conical magnetic layer
  thickness on the real (left) and imaginary (right) part of
  spin-triplet pairing correlations. The top row shows the
  unequal-spin contributions corresponding to $f_{\uparrow
    \downarrow}+f_{\downarrow \uparrow}$, whereas the lower two rows
  show contributions for $f_{\uparrow \uparrow}$ and $f_{\downarrow
    \downarrow}$, respectively. Please note the different scales
  between the top and the lower two rows.}
\end{figure}

\begin{figure}
  \centering \subfigure{
    \includegraphics[width=0.485\textwidth,clip]{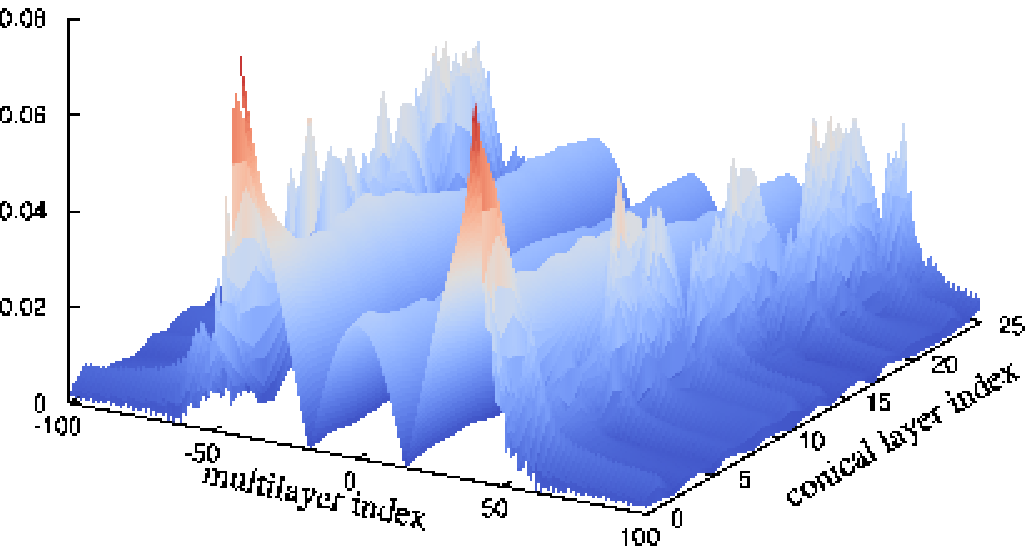}}
  \subfigure{
    \includegraphics[width=0.485\textwidth,clip]{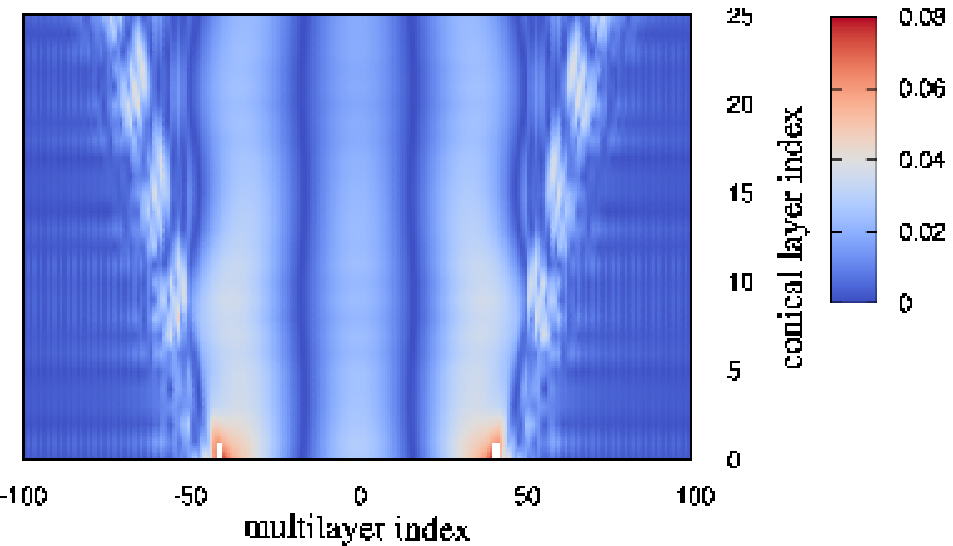}}
  \\ \subfigure{
    \includegraphics[width=0.485\textwidth,clip]{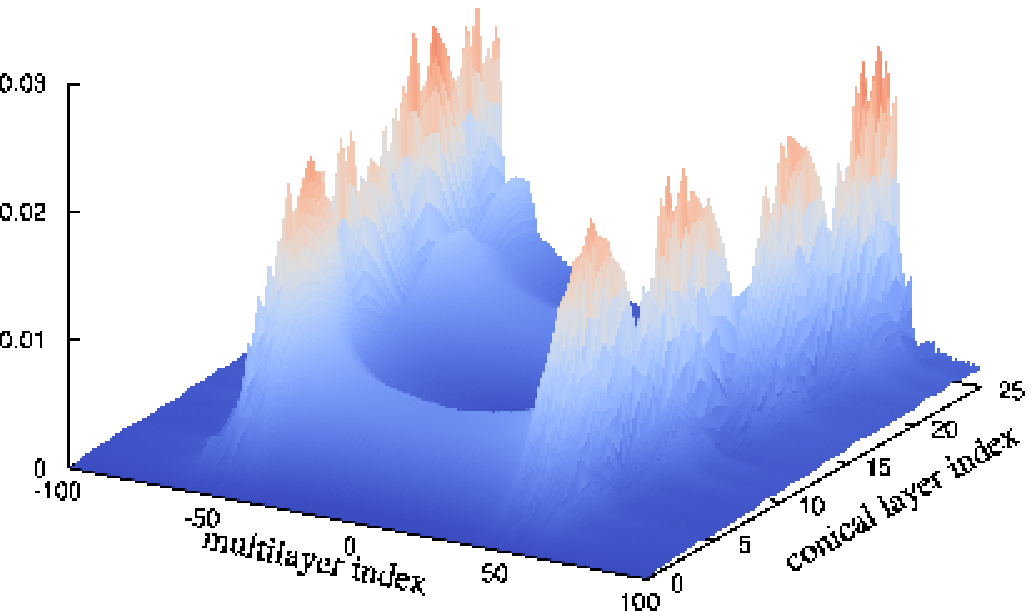}}
  \subfigure{
    \includegraphics[width=0.485\textwidth,clip]{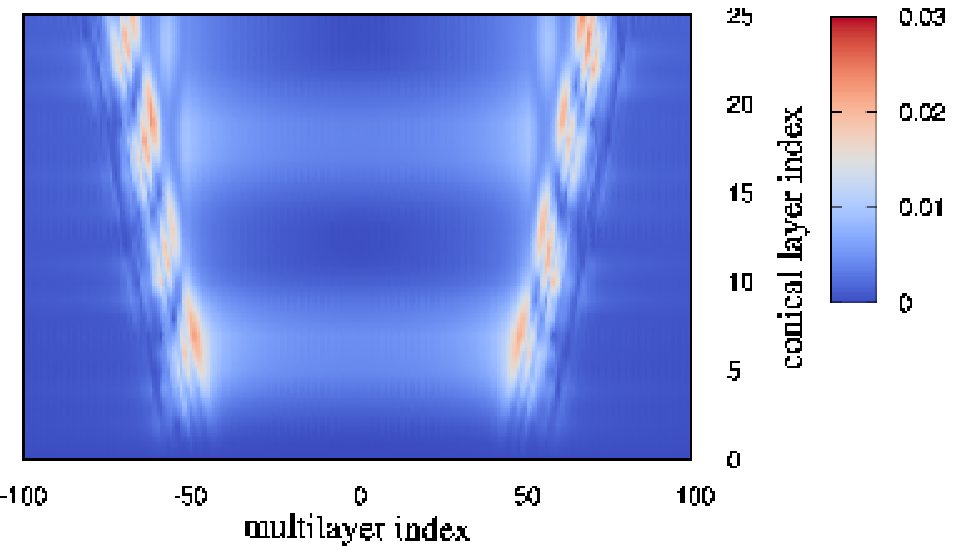}}
  \caption{\label{Fig6} Influence of increasing conical magnetic layer
    thickness on spin-triplet pairing correlations. Shown are the
    magnitude of the $f_{\uparrow \downarrow}+f_{\downarrow \uparrow}$
    (top panels) and $f_{\uparrow \uparrow}$ contributions (lower
    panels) in full view (left panels) and as a top view (right
    panels).}
\end{figure}

\begin{figure}
  \centering \subfigure{
    \includegraphics[width=0.485\textwidth,clip]{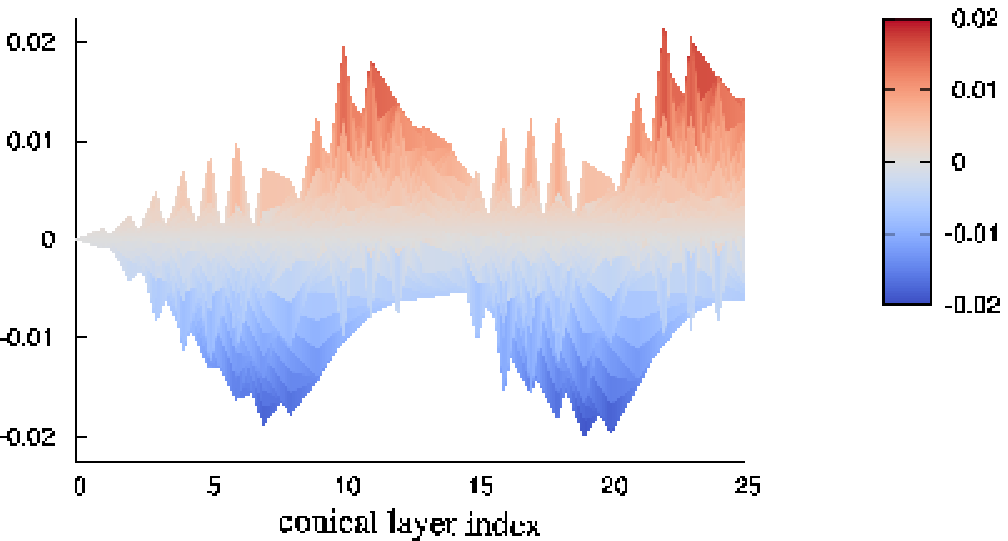}}
  \subfigure{
    \includegraphics[width=0.485\textwidth,clip]{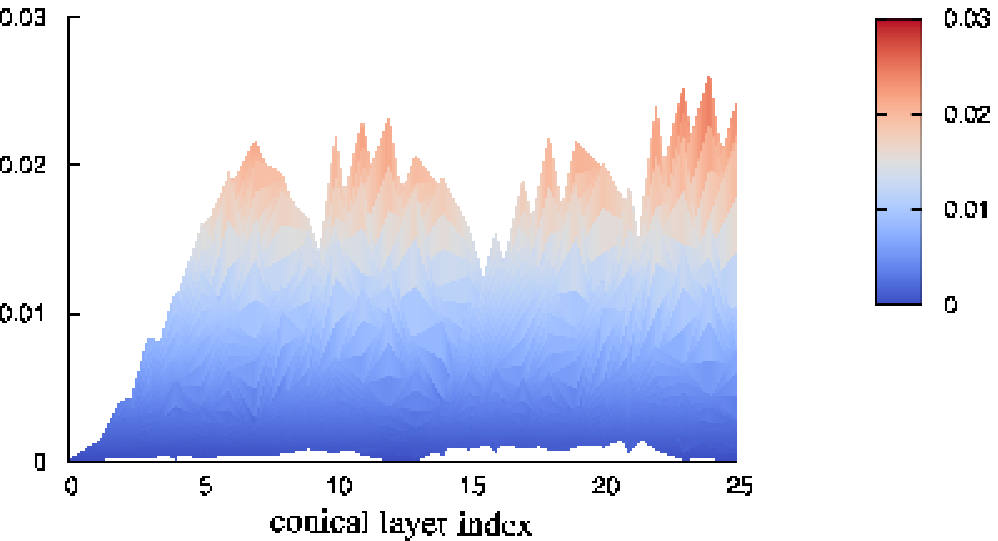}}
  \caption{\label{Fig7} Sideviews indicating a minimum number of
    conical magnetic layers to generate equal-spin spin-triplet
    pairing correlations. Shown is real part (left panel) and the
    magnitude (right panel) of the $f_{\uparrow \uparrow}$
    contribution.}
\end{figure}

The results presented in this section allow for a deeper understanding
of the influence of the conical magnetic layer thickness on the
spin-triplet pairing correlations. Since the influence of the overall
conical magnetic layer orientation and number of ferromagnetic layers
can be understood from the results already presented in \sref{Sec3_2}
and \sref{Sec3_3} the multilayer setup will now be fixed to case 1
with $n_{\rm FM} = 100$ ferromagnetic layers, but with a conical
magnetic layer thickness ranging from $n_{\rm CM} = 0$ to $n_{\rm CM}
= 25$ layers (representing two full turns of the conical magnet along
the growth direction). \Fref{Fig5} shows the influence of the conical
magnetic layer thickness on the real (left panels) and imaginary parts
(right panels) of the spin-triplet correlations for $f_{\uparrow
  \downarrow}+f_{\downarrow \uparrow}$ (upper panels), $f_{\uparrow
  \uparrow}$ (middle panels), and $f_{\downarrow \downarrow}$
contributions (lower panels), respectively.
In addition to the features already observed in the previous sections
(\fref{Fig2}) new features develop due to the increasing conical
magnetic layer thickness. For all three contributions to the
spin-triplet correlations there is an oscillating behaviour in the
real and imaginary parts depending on the conical layer thickness, in
the case of the imaginary part of the $f_{\uparrow
  \downarrow}+f_{\downarrow \uparrow}$ contribution superimposed to
the oscillations within the ferromagnetic region. Apparently, the
maximum and minimum values observed in \fref{Fig2} for the
spin-triplet correlations on either side of the ferromagnetic middle
layer are strongly affected by the conical magnetic layer thicḱness
and even change sign. However, for a fixed number $n_{\rm CM}$ the
symmetry relations observed in \sref{Sec3_1} are still valid. To get
more insight into the results \fref{Fig6} now shows the magnitudes of
the $f_{\uparrow \downarrow}+f_{\downarrow \uparrow}$ (upper panels)
and $f_{\uparrow \uparrow}$ (lower panels) spin-triplet correlations
in full view (left panels) and as a top view (right panels),
respectively.
From the magnitude of the $f_{\uparrow \downarrow}+f_{\downarrow
  \uparrow}$ spin-triplet correlation one again notices the FFLO
oscillations within the ferromagnetic region (upper panels), whereas
the lower panels clearly show that the conical magnetic layer
thickness strongly affects the strength of the $f_{\uparrow \uparrow}$
spin-triplet correlation. To get even more insight \fref{Fig7} finally
shows sideviews of the real part (left panel) and the magnitude (right
panel) of the $f_{\uparrow \uparrow}$ spin-triplet correlations. It is
apparent that a minimum number of conical magnetic layers are
necessary to generate equal-spin spin-triplet correlations, in
agreement with experimental observations. Keeping in mind that the
multilayer setup is always starting with an antiferromagnetic-like
coupling between the conical magnet and the ferromagnetic layer, an
increasing number of conical magnetic layers as displayed in the
results of \fref{Fig5} to \fref{Fig7} includes a different orientation
of the conical magnetic structure at the superconductor / conical
magnet interface. This detail requires more investigations as to
whether the specific orientation of the conical magnet at the
superconductor / conical magnet interface is partly responsible for
the oscillating behaviour shown in the results with increasing conical
magnetic layer thickness.

\section{Summary and outlook}
\label{SummaryAndOutlook}

In summary, we presented a detailed analysis of spin-triplet pairing
correlations within a superconductor/conical
magnet/ferromagnet/conical magnet/superconductor heterostructure,
similar to the ones investigated experimentally by Robinson \textit{et
  al.}~\cite{Robinson_Science329_59}. The results have been obtained
by self-consistent solutions to the microscopic spin-dependent
Bogoliubov$-$de Gennes equations which easily incorporate noncollinear
exchange fields required to model the conical magnetic structure of
Holmium also used in the experimental multilayers. While using a
similar approach as Wu \textit{et al.}~\cite{Wu_PRB86_184517} we
extended their conical magnet/superconductor bilayer investigation to
cover the whole heterostructure mentioned above. A detailed symmetry
analysis of the equal-spin spin-triplet correlations from both, the
left and the right hand side conical magnetic structure in the
heterostructure, revealed at first sight surprising relations. These
relations have been traced back to the specific underlying symmetry of
our heterostructure setup. In addition, it has been shown that, in
agreement with experimental observations, a certain minimum number of
conical magnetic layers is necessary to sufficiently generate
equal-spin spin-triplet Cooper pairs required for the long-range
triplet proximity effect.

\ack This work has been financially supported by the EPSRC
(EP/I037598/1) and made use of computational resources of the
University of Bristol. The authors gratefully acknowledge fruitful
discussion with M. Gradhand, M. G. Blamire and J. W. A. Robinson.

\end{document}